\documentclass[nofootinbib,twocolumn,aps,prl,superscriptaddress,citeautoscript,floatfix, groupedaddress]{revtex4-2}

\usepackage{graphicx}
\usepackage{amsmath,amssymb}
\usepackage[dvipsnames]{xcolor}
\usepackage{float}
\usepackage{listings}
\usepackage{mathtools}
\usepackage{titlesec}
\usepackage{romannum}

\newcommand{\beq}{\begin{eqnarray}}% can be used as {equation} or {eqnarray}
\newcommand{\eeq}{\end{eqnarray}}
\def\beqa{\begin{eqnarray}}
\def\eeqa{\end{eqnarray}}
\newcommand{\nn}{\nonumber}

\begin{document}
%%%%%%%%%
\author{Yael Avni$^1$}
\author{Ram M. Adar$^{2,3}$}
\author{David Andelman$^{1}$}
\author{Henri Orland$^{4}$}
\affiliation{${}^{1}$School of Physics and Astronomy, Tel Aviv University, Ramat Aviv 69978, Tel Aviv, Israel}
\affiliation{${}^{2}$Coll\`ege de France, 11 place Marcelin Berthelot, 75005 Paris, France}
\affiliation{${}^{3}$Institut Curie PSL University, 26 rue d'Ulm, 75248 Paris Cedex 05, France}
\affiliation{${}^{4}$Institut de Physique Th\'eorique, Universit\'e de Paris-Saclay, CEA, CNRS, F-91191 Gif-sur-Yvette Cedex, France}

%%%%%%%%%
\title{Conductivity of Concentrated Electrolytes}
%%%%%%%%%%%%%%%%%%%%%%%%%%
%%%%%%%%
\begin{abstract}
The conductivity of ionic solutions is arguably their most important trait, being widely used in electrochemical, biochemical, and environmental applications. The Debye-H\"uckel-Onsager theory successfully predicts the conductivity at very low ionic concentrations of up to a few millimolars, but there is no well-established theory applicable at higher concentrations. We study the conductivity of ionic solutions using a stochastic density functional theory, paired with a modified Coulomb interaction that accounts for the hard-core repulsion between the ions. The modified potential suppresses unphysical, short-range electrostatic interactions, which are present in the Debye-H\"uckel-Onsager theory. Our results for the conductivity show very good agreement with experimental data up to 3\,molars, without any fit parameters. We provide a compact expression for the conductivity, accompanied by a simple analytical approximation.
\end{abstract}
%%%%%%%
\maketitle

A plethora of electrokinetic phenomena occurs in electrolytes and relies on the interplay between Coulombic interactions, hydrodynamics, and thermal diffusion~\cite{Wilson2020,Alessio2021,Bazant2010,Staffan2010,Masliyah2006}. One of the most fundamental concepts in all electrokinetic phenomena is
the charge flow under an applied electric field, manifested in the electric conductivity~\cite{ElectrochemicalSystems, Bockris,Vila2006,Feng2019}.

The conductivity of electrolytes has been studied ever since the pioneering works of Debye and H\"uckel~\cite{DH} and Onsager~\cite{Onsager} in the early 20th century. They used the notion of an {\it ionic cloud}, where each ion is assumed to be surrounded by a smeared ionic distribution of net opposite charge, which gets distorted upon movement of the central ion. This led to the formulation of the reknown Debye-H\"uckel-Onsager (DHO) equation that describes the electric conductivity of ionic solutions as a function of the ion concentration~\cite{Onsager2}.

Albeit successful for dilute solutions, the DHO equation breaks down when the ion concentration exceeds the threshold of a few millimolars~\cite{Bockris}. This poses a problem as most ionic solutions in nature and in industrial applications are more concentrated~\cite{MolecularCell, Kornyshev2020, Perkin2016, Perkin2017, Turton2008}. Throughout the years, there have been many attempts to extend the DHO theory to higher concentrations, and while impressive progress has been made~\cite{OnsagerFuoss1962,Friedman1983,Chandra1999,Fraenkel2018,Bernard1991,Zhang2020}, there is still no well-established theory applicable at higher concentrations. In particular, previous works either rely on additional fit parameters that limit their predictive power, or contain exhaustive and very elaborated results that are not thoroughly transparent to the larger interdisciplinary community.

Recently, D\'emery and Dean have shown that one of the two correction terms of the DHO equation can be derived from a novel stochastic density functional theory (SDFT)~\cite{Demery2016}. Furthermore, P\'eraud~{\it et. al.}~\cite{Peraud2017,Donev2019} recovered the full DHO equation using SDFT. They included the ion advection by the fluid, which was absent in the analysis of D\'emery and Dean~\cite{Demery2016}. The SDFT analysis in Refs.~\cite{Demery2016,Peraud2017,Donev2019} is free of the notion of an ionic cloud, as it only relies on establishing the interactions between the ions, while the rest of the calculations in the dilute solution follows systematically. A natural question then arises: can SDFT be used to improve the DHO equation beyond its range of validity for high ionic concentrations?

In this Letter, we use SDFT to calculate the electric conductance of monovalent electrolytes. We introduce a simple modified interaction potential that takes into account the hard-core repulsion in an approximated manner. In addition, we subtract the self-interaction that emerges from the calculation. Our results agree well with experimental measurements up to concentrations of a few molars for different electrolytes and different temperatures, without using any adjustable parameters. Moreover, our expressions are compact, and present a clear improvement for monovalent electrolytes over previous works~\cite{Bernard1991,Chandra1999}.

{\it System description---}
We consider a monovalent ionic solution with cations and anions of charge $\pm e$, and bulk concentration $n$. The solvent is characterized by a dimensionless dielectric constant $\varepsilon$, viscosity $\eta$ and temperature $T$. The diffusion coefficient of the cations and anions at infinite ionic dilution is $D_+$ and $D_-$, respectively. The solution is subjected to an external electric field in the $\hat{x}$ direction, ${\boldsymbol{E}_{0}=E_{0}\hat{x}}$, which induces an electric current density, $J_x$, along the same direction. The conductivity of the solution is defined by the ratio
\beq \label{kaapa_0}
\kappa=\langle J_{x}\rangle/E_{0}
\eeq
where $\langle ... \rangle$ is the thermodynamic ensemble average. Although the conductivity can be calculated for any $E_{0}$, we will examine $\kappa$ in the weak-field limit, $E_0\to0$, where $\kappa$ is independent of $E_0$.

At infinite dilution ($n\to 0$), the cations and anions perform a Brownian motion with mean velocity $\pm e \mu_{\pm} E_0\hat{x}$, respectively, where $\mu_{\alpha}$ ($\alpha=\pm$) is their mobility at infinite dilution, related to the diffusion coefficient $D_{\alpha}$ by the Einstein relation, $\mu_{\alpha}=D_{\alpha}/k_{\rm B} T$, where $k_{\rm B}$ is the Boltzmann constant. The conductivity in this limit, defined as $\kappa_{0}$, is then simply
\beq \label{kappa_0}
\kappa_{0}=2 {e}^{2}\bar{\mu} n,
\eeq
where $\bar{\mu}=(\mu_+ + \mu_-)/2$ is the mean mobility. %{\YA{In practice, $\bar{\mu}$ is extracted from conductivity measurements at very small concentrations using Eq.~(\ref{kappa_0})}}.

At low concentrations, the interionic interactions reduce the conductivity. The correction to $\kappa_0$, to leading order in $n$, is given by the DHO result~\cite{Onsager2,Onsager3},
\beq\label{Onsager}
\kappa\left(n\right)=\kappa_{0}\left[1-\left(\mathcal{A}\frac{l_{\rm B}^{1/2}}{\eta\bar{\mu}}+\mathcal{B}l_{\rm B}^{3/2}\right)n^{1/2}\right],
\eeq
where $\mathcal{A}$ and $\mathcal{B}$ are numerical prefactors, $\mathcal{A}=\sqrt{2}/\left(3\sqrt{\pi}\right) \simeq 0.49$ and $\mathcal{B}=2\sqrt{\pi} \left(\sqrt{2}-1\right)/3 \simeq 0.27$, and $l_{\rm B}=e^2/(4\pi \varepsilon_0 \varepsilon k_{\rm B} T)$ is the Bjerrum length, where $\varepsilon_0$ is the vacuum permittivity.
The $\mathcal{A}$- and $\mathcal{B}$-correction terms in Eq.~(\ref{Onsager}) result from hydrodynamically mediated electrostatic interactions and direct electrostatic interactions, respectively. Traditionally, they are referred to as the {\it electrophoretic} and {\it relaxation} terms, respectively.

It is more convenient to express the conductivity in terms of physical length scales,
\beq\label{Onsager2}
\kappa\left(\lambda_{{\rm D}}\right)=\kappa_{0}\left(1-\frac{r_{\rm s}}{\lambda_{{\rm D}}}-\frac{1}{3}\left(1-\frac{1}{\sqrt{2}}\right)\frac{l_{\rm B}}{\lambda_{{\rm D}}}\right),
\eeq
where $\lambda_D=(8\pi l_{\rm B} n)^{-1/2}$ is the Debye screening length, and $r_{\rm s}=1/(6\pi\eta\bar{\mu})$ is a reduced Stokes radius, different from the physical ion radii. For simple aqueous solutions at room temperature, typical length scales are ${l_{\rm B} \sim 7\,{\rm \AA}}$, ${r_{\rm s} \sim 1\, {\rm \AA}}$, and ${\lambda_{\rm D}\sim 3 [{\rm \AA}]/\sqrt{n \rm [M]}}$.
Note that the most pronounced deficiency of the DHO equation is that it accounts for electrostatic attraction between oppositely charged ions at unrealistic distances: smaller than the ionic size (see Fig.~\ref{Fig1}).

%%%%%%%%%%%%%%%%%%%%%%%%%%%%%%%%%%%%%%%%%%%
%Fig1
\begin{figure}
\includegraphics[width = 0.8 \columnwidth,draft=false]{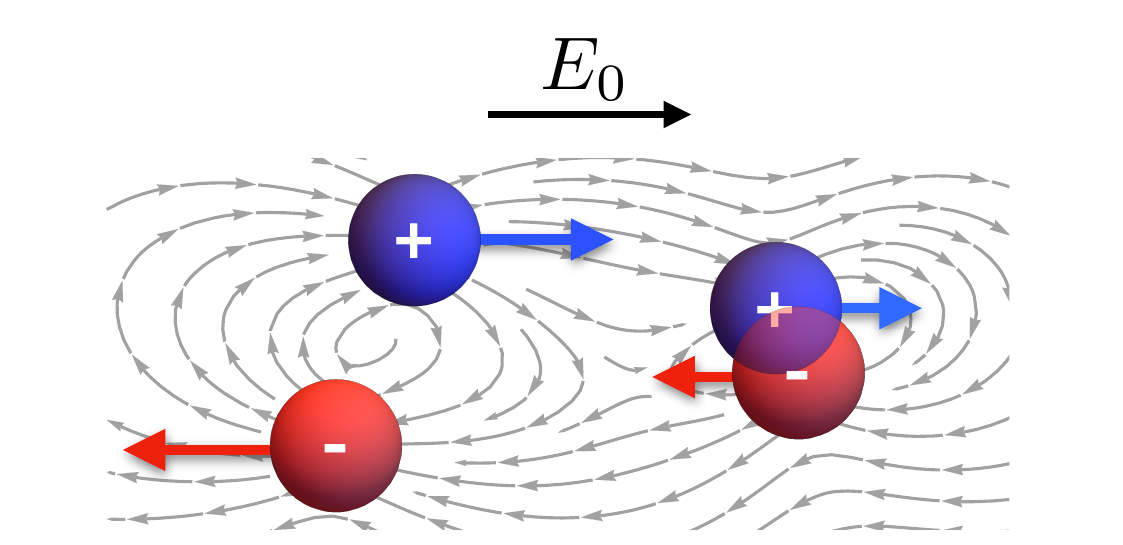} %0.4
\caption{\textsf{A schematic drawing of cations (blue) and anions (red) moving in response to an applied electric field $\boldsymbol{E_0}$. The grey lines represent the fluid velocity field. If the interaction is purely Coulombic, oppositely charged ions are likely to get unrealistically close to one another (right side), thus reducing the conductivity. We use a modified potential to avoid such proximity, prohibited by the ionic finite size.}}
\label{Fig1}
\end{figure}
%%%%%%%%%%%%%%%%%%%%%%%%%%%%%%%%%%%%%%%

{\it The equations of motion ---}
We denote the local ionic concentrations by $n_{\pm}(\boldsymbol{ r})$. Since the number of particles is conserved, the ionic concentrations satisfy the continuity equation
\beq \label{continuity}
\partial_{t}n_{\alpha} = -\boldsymbol{\nabla}\cdot\boldsymbol{j}_{\alpha}\,\,\,\,\,\,\,\,\,\,\,\, \alpha={\pm},
\eeq
where $j_{\pm}$ are the positive and negative ionic fluxes. For dilute solutions, the fluxes are given by
\beq \label{j}
\boldsymbol{j}_{\alpha}=n_{\alpha}\boldsymbol{u}-D_{\alpha}\boldsymbol{\nabla}n_{\alpha}+\mu_{\alpha}\boldsymbol{f}_{\alpha}-\sqrt{2D_{\alpha}n_{\alpha}}\boldsymbol{\zeta}_{\alpha},
\eeq
where $\boldsymbol{u}$ is the solvent velocity field, $\boldsymbol{f}_{\pm}$ are the electrostatic force densities exerted on the cations and anions, respectively, and $\boldsymbol{\zeta}_{\pm}$ are 3D white noise functions, satisfying
\beqa \label{noise}
\langle\boldsymbol{\zeta}_{\alpha}\left(\boldsymbol{r},t\right)\rangle & = & 0,\\
\langle{\zeta}_{\alpha}^{n}\left(\boldsymbol{r},t\right){\zeta}_{\beta}^{m}\left(\boldsymbol{r}',t'\right)\rangle & = & \delta_{\alpha \beta}\delta_{nm}\delta\left(t-t'\right)\delta\left(\boldsymbol{r}-\boldsymbol{r}'\right), \nn
\eeqa
where $n$ and $m$ denote the Cartesian components.
The first term in Eq.~(\ref{j}) is the ionic advection by the solvent, while the second and third terms constitute the electrochemical potential gradient, which acts as a driving force. The last term is a stochastic flux, present in the SDFT formalism, which can be derived by different means such as writing the Langevin equation in terms of the local ionic concentrations~\cite{Kawasaki1993,Dean1996,DeGroot,Basu2008}. The stochastic flux is responsible for the dynamics beyond mean field, and its coefficient, $\sqrt{2D_{\alpha}n_{\alpha}}$, guarantees that the fluctuation-dissipation theorem is satisfied.

The force densities $\boldsymbol{f}_{\pm}$ originate from the external force generated by the electric field $\boldsymbol{E}_{0}$ and the interactions between the ions. It can be written as
\beq \label{force_density}
\boldsymbol{f}_{\alpha}=n_{\alpha}e_{\alpha}\boldsymbol{E}_{0}-n_{\alpha}\sum_{\beta}\int{\rm d}^{3}r'n_{\beta}\left(\boldsymbol{r}'\right)\boldsymbol{\nabla}v_{\alpha \beta}\left(\left|\boldsymbol{r}-\boldsymbol{r}'\right|\right)\,\,\,\,\,
\eeq
where $v_{\alpha \beta}$ is the pair-interaction energy between ions of species $\alpha$ and $\beta$. For the standard Coulomb interaction, $v_{\alpha \beta}\left(r\right)=e_{\alpha} e_{\beta}/(4\pi\varepsilon_{0}\varepsilon r)$, where $e_{\alpha}$ and $e_{\beta}$ are the charges ($\pm e$). However, the finite ion size prevents ions from getting very close to one another (see Fig.~\ref{Fig1}). This can be described, to a good approximation, by a hard-core potential, namely, taking ${v_{\alpha \beta}(r<a) \to \infty}$, where $a$ is a cutoff length (distance of closest approach) that we identify as the sum of the cation and anion physical radii, ${a=r_+ + r_-}$. It is not possible to include such a diverging interaction within our perturbative approach. Instead, a viable modification is to apply a cutoff to the Coulomb interaction~\cite{Adar2019},
\beqa \label{u_co}
v_{\alpha \beta}\left(r\right)=s_{\alpha}s_{\beta}V_{\text{co}}\left(r\right)\nn \\
V_{\text{co}}\left(r\right)\equiv\frac{{e}^{2}}{4\pi\varepsilon_{0}\varepsilon r}\theta\left(r-a\right),
\eeqa
where $s_{\pm}=\pm1$ and $\theta(r)$ is the Heaviside function.
Equation~(\ref{u_co}) does not contain a hard-core repulsion that prohibits the ions from overlapping. However, as we show in the Supplemental Material~\cite{Supplemental} for a simplified system, Eq.~(\ref{u_co}) approximates very well (far better than the pure Coulomb interaction) the average distance between oppositely charged ions in a Coulomb gas with hard-core repulsion, for concentrations up to a few molars. For that reason, we will use it hereafter. Inserting the new interaction potential, the ionic fluxes are
\beqa \label{fluxes}
\boldsymbol{j}_{\alpha}= && n_{\alpha}\boldsymbol{u}-D_{\alpha}\nabla n_{\alpha}+\mu_{\alpha}n_{\alpha}e_{\alpha}\boldsymbol{E}_{0}-\sqrt{2D_{\alpha}n_{\alpha}}\boldsymbol{\zeta}_{\alpha}\nn\\
 && -\mu_{\alpha}n_{\alpha}s_{\alpha}\int{\rm d}^{3}r'\rho\left(\boldsymbol{r}'\right)\boldsymbol{\nabla}V_{\text{co}}\left(\left|\boldsymbol{r}-\boldsymbol{r}'\right|\right)
\eeqa
where we defined the concentration difference ${\rho\left(\boldsymbol{r}\right)\equiv n_{+}\left(\boldsymbol{r}\right)-n_{-}\left(\boldsymbol{r}\right)}$.

Last, the solvent velocity $\boldsymbol{u}$ satisfies the incompressibility condition,
\beq \label{incompres}
\boldsymbol{\nabla}\cdot \boldsymbol{u}=0,
\eeq
and the Stokes equation is given by
\beq \label{Stokes}
\eta\nabla^{2}\boldsymbol{u}-\boldsymbol{\nabla}p+\boldsymbol{f}_{+}+\boldsymbol{f}_{-}=0,
\eeq
where $p$ is the pressure and the drag force exerted on the solvent by the ions was equated to the electrostatic force acting on the ions. Substituting the interaction potential in the expressions for $\boldsymbol{f}_{\pm}$, the last equation becomes
\beqa \label{NS1}
\eta\nabla^{2}\boldsymbol{u} && =\boldsymbol{\nabla}p-e\rho\left(\boldsymbol{r}\right)\boldsymbol{E}_{0}\\
 && +\,\rho\left(\boldsymbol{r}\right)\int{\rm d}^{3}r'\rho\left(\boldsymbol{r}'\right)\boldsymbol{\nabla}V_{\text{co}}\left(\left|\boldsymbol{r}-\boldsymbol{r}'\right|\right).\nn
\eeqa
Equations~(\ref{continuity}),~(\ref{fluxes}),~(\ref{incompres}), and~(\ref{NS1}) govern the dynamics and determine the conductivity.

{\it Calculation of the conductivity --}
To calculate the conductivity, $\kappa=\langle J_x \rangle/E_0$, we recall that the current density $\boldsymbol J$ is related to the ionic fluxes by $\boldsymbol{J}=e\left(\boldsymbol{j}_{+}-\boldsymbol{j}_{-}\right)$, where $\boldsymbol{j}_{\pm}$ are given by Eq.~(\ref{fluxes}). As the system is homogeneous, the local ionic concentrations satisfy $\langle n_{\pm}({\boldsymbol r})\rangle =n\nn$, resulting in the following expression for the conductivity:
\beq \label{Full_kappa}
\kappa=\kappa_{0}+\kappa_{\text{hyd}}+\kappa_{\text{el}},
\eeq
where
\beq \label{kappa_hyd}
\kappa_{\text{hyd}}=\frac{e}{E_{0}}\langle u_{x}\left(\boldsymbol{r}\right)\rho\left(\boldsymbol{r}\right)\rangle
\eeq
and
\beqa \label{kappa_ele}
\kappa_{\text{el}}=-\sum_{\alpha=\pm}\frac{e\mu_{\alpha}}{E_{0}}\langle n_{\alpha}\left(\boldsymbol{r}\right)\int{\rm d}^{3}{r}'\rho\left(\boldsymbol{r}'\right)\partial_{x}V_{\text{co}}\left(\left|\boldsymbol{r}-\boldsymbol{r}'\right|\right)\rangle.\,\,\,\,\,\,\,\,
\eeqa
The first correction to $\kappa_0$, $\kappa_{\text{hyd}}$, represents the hydrodynamically mediated electrostatic interactions (electrophoretic term). The second correction, $\kappa_{\text{el}}$, results from direct electrostatic interactions (relaxation term), yet it incorporates intrinsically the hard-core repulsion, through the short-distance cutoff of $V_{\text{co}}$. The average in Eq.~(\ref{kappa_hyd}) includes the ion self-interaction that should be subtracted, as we will do later on.

The averages in Eq.~(\ref{kappa_hyd}) and Eq.~(\ref{kappa_ele}) cannot be done exactly. Instead, we linearize the equations of motion~\cite{Demery2016,Peraud2017}. We write $n_{\pm}({\boldsymbol r})  =n+\delta n_{\pm}({\boldsymbol r})$, $\rho ({\boldsymbol r})  =\delta\rho({\boldsymbol r})$, $\boldsymbol{u}({\boldsymbol r})  =\delta\boldsymbol{u}({\boldsymbol r})$ and $p({\boldsymbol r})  =p_0 + \delta{p}({\boldsymbol r})$,
and keep only terms up to linear order in $\delta n_{\pm}$, $\delta\rho$, $\delta \boldsymbol{u}$, $\delta p$, and $\zeta_{\pm}$. Defining for any function $f(\boldsymbol{r})$ its Fourier transform $\tilde{f}(\boldsymbol{k})=\int {\rm d}^3r f(\boldsymbol{r}){\rm e}^{-i\boldsymbol{k}\cdot\boldsymbol{r}}$, the linearized equations can be written in a simple matrix form in Fourier space (derivation is given in the Supplemental Material~\cite{Supplemental}),
\beq \label{matrix_equation}
\frac{\partial \delta{\tilde{n}}_{\alpha}(\boldsymbol{k})}{\partial t}=A_{\alpha \beta}(\boldsymbol{k})\delta\tilde{n}_{\beta}(\boldsymbol{k})+B_{\alpha \beta}(\boldsymbol{k})\tilde{\zeta}_{\beta}(\boldsymbol{k}).
\eeq
where $A(\boldsymbol{k})$ and $B(\boldsymbol{k})$ are the matrices
\beqa
A_{\alpha \beta}(\boldsymbol{k}) && =\begin{cases}
-D_{\alpha}k^{2}-\mu_{\alpha}nk^{2}\tilde{V}_{\text{co}}(\boldsymbol{k}) -i\mu_{\alpha}e_{\alpha}k_{x}E_{0} & \alpha=\beta\nn\\[3pt]
\mu_{\alpha}nk^{2}\tilde{V}_{\text{co}}(\boldsymbol{k}) & \alpha\neq \beta
\end{cases}\\
B_{\alpha \beta}(\boldsymbol{k})&&=i\sqrt{2D_{\alpha} n}k\delta_{\alpha \beta},
\eeqa
and $\tilde{\zeta}_{\pm}(\boldsymbol{k})$ are white-noise scalar functions: $\langle\tilde{\zeta}_{\alpha}(\boldsymbol{k})\rangle =0$, ${\langle\tilde{\zeta}_{\alpha}(\boldsymbol{k})\tilde{\zeta}_{\beta}(\boldsymbol{k}')\rangle =\left(2\pi\right)^{3}\delta_{\alpha \beta}\delta\left(t-t'\right)\delta\left(\boldsymbol{k}+\boldsymbol{k}'\right)}$.

%%%%%%%%%%%%%%%%%%%%%%%%%%%%%%%%%%%%%%%%%%%
%Fig2
\begin{figure*}
\includegraphics[width = 2 \columnwidth,draft=false]{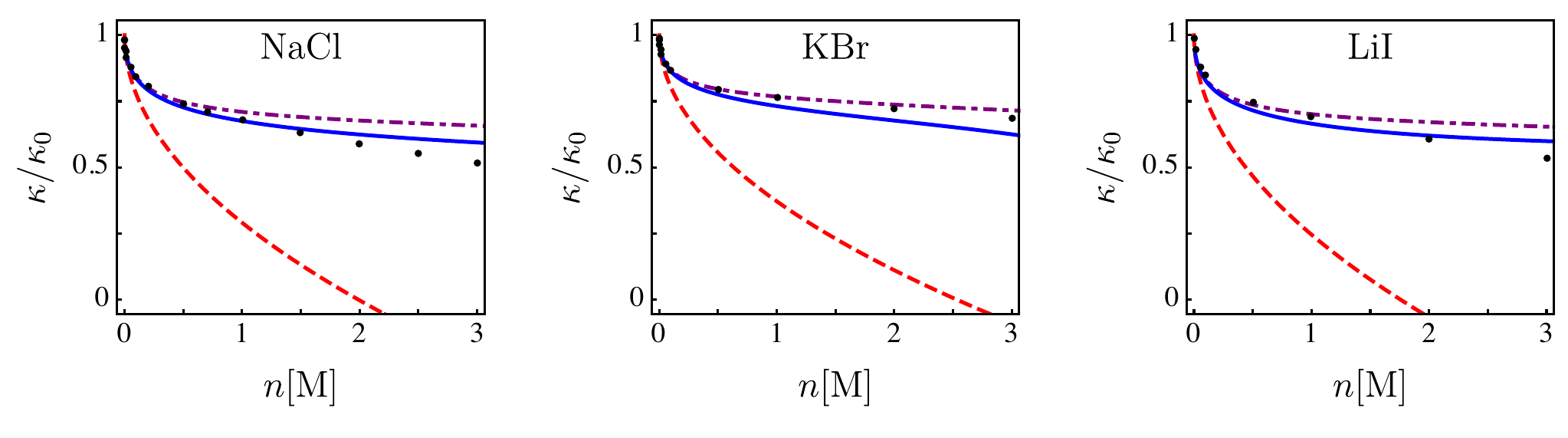} %0.4
\caption{\textsf{The conductivity, $\kappa$, of (a) NaCl, (b) KBr, and (c) LiI in water at $T=25^{\circ}$C, normalized by $\kappa_0$, as a function of the salt concentration $n$. Black dots - experimental data~\cite{Lide_Book,Lobo}; full blue line - numerical result, Eqs.~(\ref{result1}) and~(\ref{result2}); dotted-dashed purple line - analytical approximation, Eq.~(\ref{approx}); dashed red line - the DHO theory. Radii from crystallographic data: $r_{\text{Na}}=1.02\,{\rm \AA}$, $r_{\text{Cl}}=1.81\,{\rm \AA}$, $r_{\text{K}}=1.38\,{\rm \AA}$, $r_{\text{Br}}=1.96\,{\rm \AA}$, $r_{\text{Li}}=0.76\,{\rm \AA}$, $r_{\text{I}}=2.2\,{\rm \AA}$. Other physical parameters are specified in the Supplemental Material~\cite{Supplemental}.
}}
\label{Fig2}
\end{figure*}
%%%%%%%%%%%%%%%%%%%%%%%%%%%%%%%%%%%%%%%
%%%%%%%%%%%%%%%%%%%%%%%%%%%%%%%%%%%%%%%%%%%
%Fig3
\begin{figure}
\includegraphics[width = 0.9 \columnwidth,draft=false]{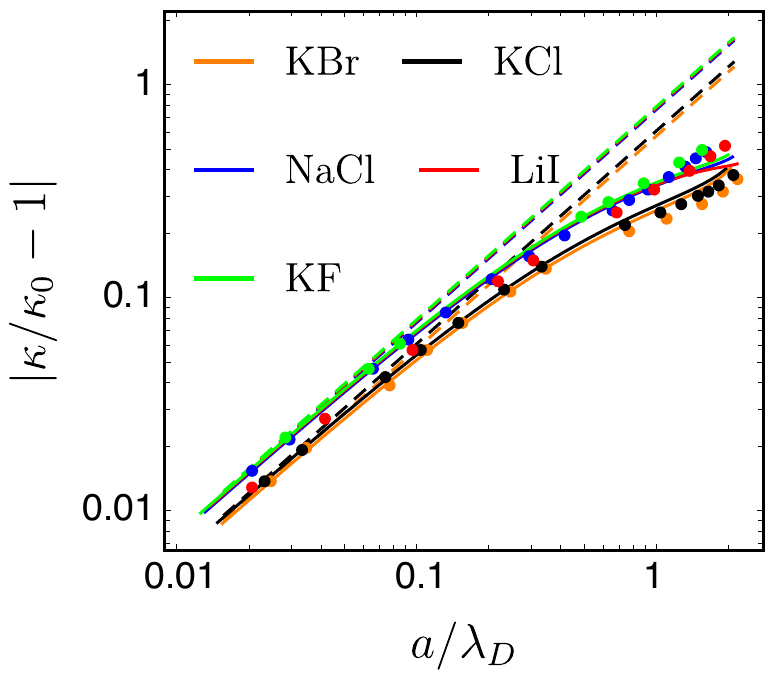} %0.4
\caption{\textsf{The relative conductivity correction, $|\kappa/\kappa_0-1|$, as a function of $a/\lambda_{\rm D}$, for different salts on a log-log plot. Dots - experimental data~\cite{Lide_Book,Lobo}; Full lines - numerical result, Eqs.~(\ref{result1}) and~(\ref{result2}); Dashed lines - the DHO theory. Note that the blue, red and green theoretical lines, corresponding to NaCl, LiI and KF, respectively, are almost indistinguishable in the figure. Physical parameters are the same as in Fig.~\ref{Fig2}, with the additional ionic radius: $r_{\rm{F}}=1.33\,{\rm \AA}$.}}
\label{Fig3}
\end{figure}
%%%%%%%%%%%%%%%%%%%%%%%%%%%%%%%%%%%%%%%

For the linear system of equations in Eq.~(\ref{matrix_equation}), it can be shown that
\beq \label{correlator_k}
\langle\delta\tilde{n}_{\alpha}(\boldsymbol{k})\delta\tilde{n}_{\beta}(\boldsymbol{k'})\rangle=\left(2\pi\right)^{3}C_{\alpha \beta}(\boldsymbol{k})\delta\left(\boldsymbol{k}+\boldsymbol{k}'\right),
\eeq
where the correlation matrix, $C(\boldsymbol{k})$, is given by the equation~\cite{RobertZwanzig}
\beq \label{mat_eq}
A(\boldsymbol{k})C(\boldsymbol{k})+C(\boldsymbol{k})A^{\dagger}(\boldsymbol{k})=-B(\boldsymbol{k})B^{\dagger}(\boldsymbol{k}),
\eeq
where $\dagger$ represents the Hermitian conjugate.
The conductivity correction terms, $\kappa_{\text{hyd}}$ and $\kappa_{\text{el}}$, can now be written as~\cite{Supplemental},
\beqa \label{integrals}
\kappa_{{\rm hyd}}&&=\frac{2{e}^{2}}{\eta}\int\frac{{\rm d}^{3}k}{\left(2\pi\right)^{3}}\frac{1}{k^{2}}\left(1-\frac{k_{x}^{2}}{k^{2}}\right)\big(C_{++}(\boldsymbol{k})-\text{Re}\left[C_{+-}(\boldsymbol{k})\right]\big)
\nonumber \\
\kappa_{{\rm el}}&&=\frac{2\bar{\mu}e}{E_{0}}\int\frac{{\rm d}^{3}k}{\left(2\pi\right)^{3}}\:k_{x}\tilde{V}_{\text{co}}({k})\text{Im}\left[C_{+-}(\boldsymbol{k})\right].
\eeqa

The hydrodynamic correction to the conductivity, $\kappa_{\text{hyd}}$, depends on the correlator ${C_{\alpha\alpha}(\boldsymbol{k}) \propto \langle \delta\tilde{n}_{\alpha}(\boldsymbol{k})\delta\tilde{n}_{\alpha}(\boldsymbol{k'})\rangle}$. This term includes the ion self-interaction that should be subtracted. This is done by taking the following renormalization of the correlation matrix~\cite{Supplemental},
\beq \label{norm}
C_{\alpha \beta}(\boldsymbol{k}) \to C_{\alpha \beta}(\boldsymbol{k})-n \delta_{\alpha \beta}.
\eeq
Finally, the correlation matrix is obtained by solving Eq.~(\ref{mat_eq}) and applying the normalization in Eq.~(\ref{norm}). Exact expressions for $C_{\alpha \beta}(\boldsymbol{k})$ are given in the Supplemental Material~\cite{Supplemental}.

{\it Results and comparison with experiments ---}
Substituting $C_{\alpha \beta}(\boldsymbol{k})$ in Eq.~(\ref{integrals}), taking the $E_0 \to 0$ limit, and performing the angular part of the integrals in $k$ space, we obtain
\beqa \label{result1}
\kappa_{{\rm hyd}}=-\frac{2}{\pi}\frac{\kappa_{0} r_{\rm s}}{\lambda_{{\rm D}}}\int\limits _{0}^{\infty}{\rm d}x\frac{\cos\left(\frac{ax}{\lambda_{{\rm D}}}\right)}{\cos\left(\frac{ax}{\lambda_{{\rm D}}}\right)+x^{2}}
\eeq
and
\beqa  \label{result2}
\\
\kappa_{{\rm el}}=-\frac{1}{3\pi}\frac{\kappa_{0}l_{\rm B}}{\lambda_{{\rm D}}}\int\limits _{0}^{\infty}{\rm d}x\frac{x^{2}\cos^{2}\left(\frac{ax}{\lambda_{{\rm D}}}\right)}{x^{4}+\frac{3}{2}x^{2}\cos\left(\frac{ax}{\lambda_{{\rm D}}}\right)+\frac{1}{2}\cos^{2}\left(\frac{ax}{\lambda_{{\rm D}}}\right)}\nn
\eeq
where we used the change of variables $x = \lambda_D k$.
Together with the definition, $\kappa=\kappa_{0}+\kappa_{\text{hyd}}+\kappa_{\text{el}}$ , Eqs.~(\ref{result1})-(\ref{result2}) are our main results.

While the integrals in Eqs.~(\ref{result1}) and (\ref{result2}) cannot be performed analytically, they can be approximated. To leading order in $a/\lambda_D $, we can replace $\cos(a x/\lambda_{{\rm D}})$ by unity in the denominator.
The integrals can then be evaluated using the residue theorem in the complex plane, yielding
\beqa \label{approx}
\kappa\left(\lambda_{{\rm D}}\right) && \approx\kappa_{0}\bigg(1-\frac{r_{\rm s}}{\lambda_{{\rm D}}}{\rm e}^{-a/\lambda_{{\rm D}}}\\
 && -\frac{1}{6}\left(1-\frac{1}{\sqrt{2}}+{\rm e}^{-2a/\lambda_{{\rm D}}}-\frac{1}{\sqrt{2}}{\rm e}^{-\sqrt{2}a/\lambda_{{\rm D}}}\right)\frac{l_{\rm B}}{\lambda_{{\rm D}}}\bigg). \nn
\eeqa
Equation~(\ref{approx}) recovers the DHO equation in the $a \ll \lambda_{\rm D}$ limit. As the concentration increases (and $\lambda_{\rm D}$ decreases), it predicts a larger conductivity compared to the DHO equation. This is because the finite ion-size limits the strength of the electrostatic attraction between oppositely charged ions, and this strength is responsible for reducing the conductivity at high concentrations.

In Fig.~\ref{Fig2},  the numerical evaluations of the integrals are compared with experimental data for three standard salts NaCl, KBr, and LiI in water solutions for concentrations up to $3$\,M. The experimental data are taken from Refs.~\cite{Lide_Book} and~\cite{Lobo}, where an extensive dataset of measurements is summarized. For each solution, we use the relation, $a=r_+ + r_-$, and take the ion radii extracted from crystallographic data, without any fit parameters. The exact values of the physical parameters are given in the Supplemental Material~\cite{Supplemental}. Up to $1$\,M concentrations, the agreement is excellent in all cases. Surprisingly, the agreement even at concentrations as high as $3$\,M still works very well, with the largest deviation being~$14\%$ for NaCl at $3$\,M. This is quite remarkable since the solution is no longer dilute at such high concentrations. Moreover, the physical solvent parameters such as the permittivity $\varepsilon$ are no longer constant~\cite{Adar2018}. The analytical approximation, Eq.~(\ref{approx}), is also shown in Fig.~\ref{Fig2}. It predicts slightly higher conductivities than the numerical expressions, yet it works very well, especially for KCl. Our numerical results for five different salts in water are presented on a master plot in Fig.~\ref{Fig3}. The relative conductivity correction, $|\kappa/\kappa_0-1|$ is shown as a function of $a/\lambda_{\rm D}$, which is a natural parameter as evident from Eq.~(\ref{approx}).

\begin{figure}
\includegraphics[width = 0.85 \columnwidth,draft=false]{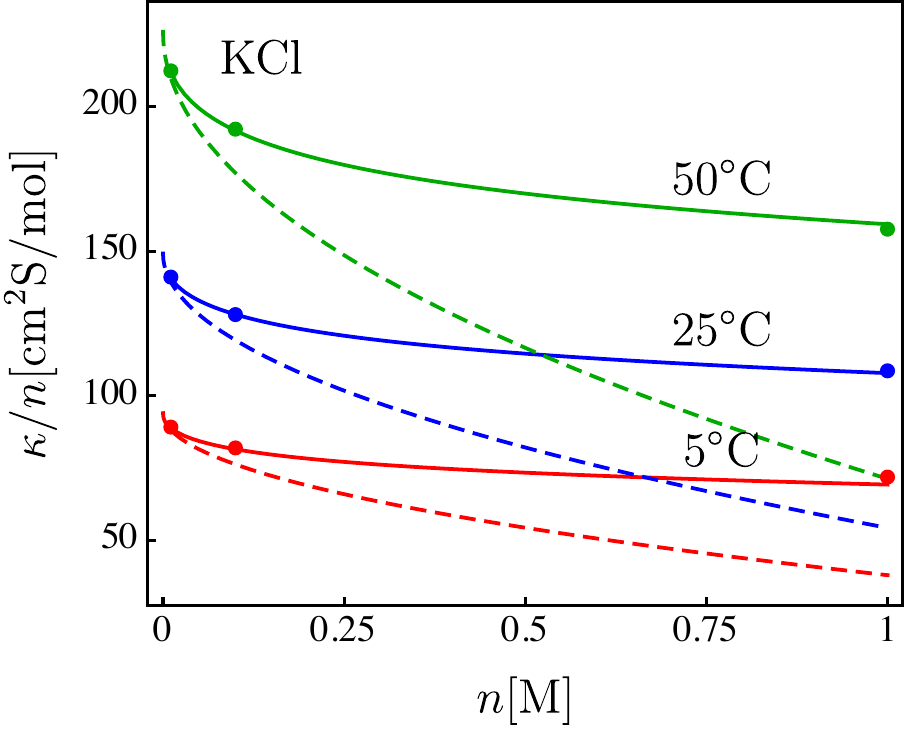} %0.4
\caption{\textsf{The conductivity, $\kappa$, of KCl, normalized by the salt concentration $n$ (S is the Siemens electric conductance unit), as a function of $n$, at temperature $T=5^\circ$C (red), $T=25^\circ$C (blue), $T=50^\circ$C (green). Experimental data is represented by dots, numerical result is in full lines, and the DHO theory is shown in dashed lines. For each temperature, $\kappa_{0}$ was set by equating the value of $\kappa$ to the experimental data at $n=0.01$\,M. Other physical parameters are specified in the Supplemental Material~\cite{Supplemental}.}}
\label{Fig4}
\end{figure}

In Fig.~\ref{Fig4}, we compare our numerical results to experimental data of KCl at three different temperatures: $5^\circ$C, $25^\circ$C and $50^\circ$C. For $5^\circ$C and $50^\circ$C data are available only in the range $0.01<n<1$\,M. Thus, we find $\kappa_0$ by equating $\kappa$ to the experimental conductivity at $0.01$\,M (we do so for $25^\circ$C as well, for consistency in the plot). Our results are very accurate for these three temperatures up to $1$\,M.

%%%%%%%%%%%%%%%%%%%%%%%%%%%%%%%%%%%%%%%%%%%
%Fig4
%%%%%%%%%%%%%%%%%%%%%%%%%%%%%%%%%%%%%%%

In conclusion, we calculated the electric conductivity of electrolytes containing monovalent ions, using the stochastic density functional theory. We account for essential finite-size effects, which are missing in the Debye-H\"uckel-Onsager theory, by introducing a modified Coulomb potential that suppresses unphysical, short-range electrostatic interactions. Our results are in excellent agreement with experimental data and provide a simple expression for the conductivity at concentrations as high as 3\,M. The theory can be generalized to multicomponent electrolytes and multivalent ions, albeit the latter is expected to limit the validity of the theory to lower concentrations due to strong electrostatic interactions. Finally, our results support SDFT as a useful tool to solve complex transport phenomena.

%\bigskip\bigskip
\acknowledgments
We would like to thank V. D\'emery, H. Diamant, A. Donev, Y. Kantor, A. Kornyshev, K. Mallick, R. Netz, P. Pincus, S. Safran and H. Stone for fruitful discussions and correspondence. Y. A. is thankful for the support of the Clore Scholars Programme of the Clore Israel Foundation. R. M. A. acknowledges support
by the Rothschild Fellowship and FRM Postdoctoral Fellowship. This work was supported by the Israel Science Foundation (ISF) under Grant No. 213/19 and by the National Natural Science Foundation of China (NSFC) -- ISF joint program under Grant No. 3396/19.

 %%%%%%%%%%%%%%%%%%%%%%%%%%%%%%%%%%

\onecolumngrid
\appendix
\section*{The Conductivity of Concentrated Electrolytes: Supplemental Material}

\section{\Romannum{1}. Testing the modified interaction potential}
\label{Testing}
%%%%%%%%%
Our main results are based on the use of a modified cutoff potential, Eq.~(9) in the Letter. The purpose of the modified potential is to account for the hard-core repulsion between the ions in a way that does not break the perturbative calculation (as do conventional hard-core potentials).
\\[3pt]
\indent To test whether the cutoff potential correctly mimics a hard-core potential, we look at a simple equilibrium variable: the average distance between two ions, having charge $e_1$ and $e_2$, respectively, placed inside a sphere of radius $R$, where one of the ions is fixed at the center, while the other is free to move inside the sphere. This is similar to dividing our ionic solutions into cells, where each cell contains two ions that do not interact with other cells, and the ion concentration $n^*$ is given by $(n^{*})^{-1}\approx (4\pi/3)R^{3}$. We denote the ion concentration with a star as $n^*$ is an approximation of the ion concentration of the real system, $n$, but it does not equal it. 
\\[3pt]
\indent
We consider three cases for the ion interaction:

\begin{itemize}
\item Coulomb interaction,
\beqa \label{u_hc}
v_{\text{c}}\left(r\right)=
\frac{e_{i} e_{j}}{4\pi\varepsilon_{0}\varepsilon r},
\eeqa
with $i,j=1,2$ for the two ions, 
\item Coulomb interaction with a hard core interaction,
\beqa \label{u_hc}
v_{\text{c}}^{\text{hc}}\left(r\right)=\begin{dcases}
\frac{e_{i} e_{j}}{4\pi\varepsilon_{0}\varepsilon r} & r>a\\
\infty & r<a,
\end{dcases}
\eeqa
\item the modified cutoff potential $v_{\text{co}}$,
\beqa \label{u_hc}
v_{\text{co}} = \frac{e_{i} e_{j}}{4\pi\varepsilon_{0}\varepsilon r} \theta\left({r-a}\right).
\eeqa
where $\theta(r)$ is the Heaviside function, and $V_{\text{co}}$, which is used in the Letter, is the absolute value of $v_{\text{co}}$. 
\end{itemize}

The second interaction, $v_{\text{c}}^{\text{hc}}$, is considered the most accurate one, although it cannot be used for our purposes. The average distance between the particles, $\langle r \rangle$, for a general interaction potential $v(r)$, is
\beqa
\langle r\rangle =\frac{\int\limits _{0}^{R}{\rm d}r\,r^{3}{\rm e}^{-\beta v\left(r\right)}}{\int\limits _{0}^{R}{\rm d}r\,r^{2}{\rm e}^{-\beta v\left(r\right)}}.
\eeqa

In Fig.~\ref{Fig1}, we plot $\langle r \rangle$ as a function of the concentration $n^*$, both for equal charges and opposite charges. The figure shows that $v_{\text{co}}$ and $v_{\text{c}}^{\text{hc}}$ produce very similar average distances, up to $n^*=3$\,M. This is the case for both equally charged particles and oppositely charged ones, although the agreement is better for the latter. The pure Coulomb interaction, while capturing very well the average distance between equal charges, completely misses the opposite-charge case. Due to the infinite attraction at short distances, it predicts $\langle r \rangle = 0$ instead of a finite value. We note that there is a small yet visible difference between $v_{\text{co}}$ and $v_{\text{c}}^{\text{hc}}$ for equal charges, at concentrations above $n^*=0.5$\,M. This explains, at least partially, why our results are less accurate in this regime (see Fig.~2 of the Letter).

%%%%%%%%%%%%%%%%%%%%%%%%%%%%%%%%%%%%%%%%%%%
%Fig1
\begin{figure}
\includegraphics[width = 0.9 \columnwidth,draft=false]{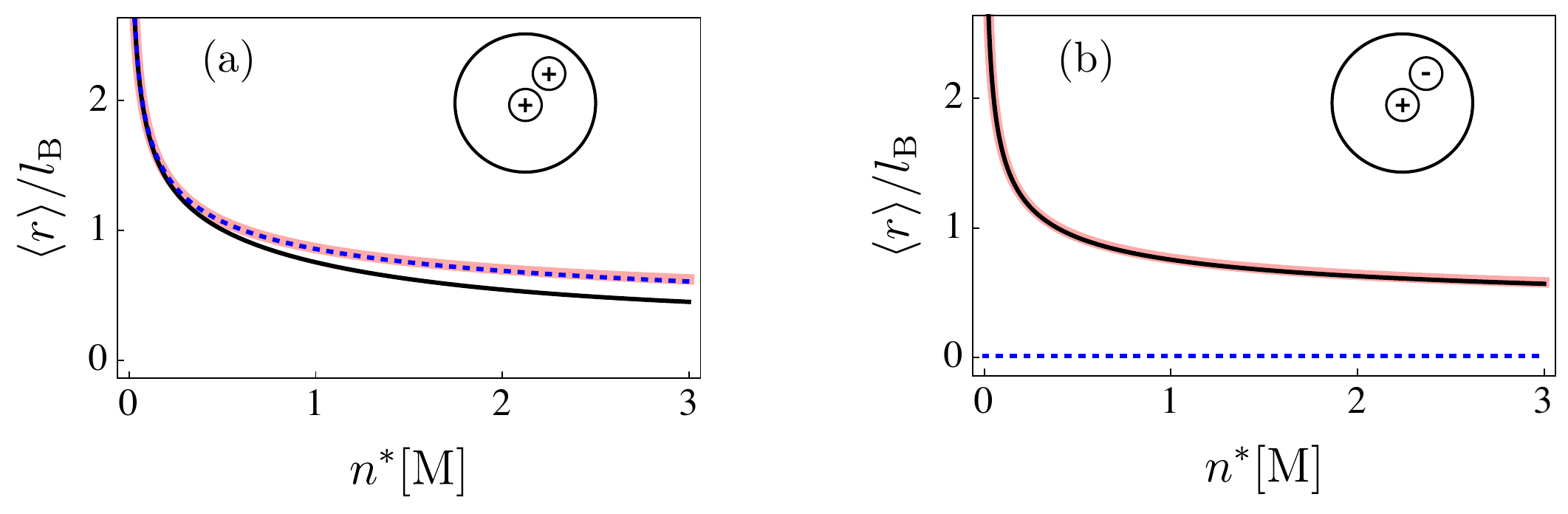} %0.4
\caption{\textsf{Average distance between (a) two equally and (b) two oppositely charged ions, as a function of the ion concentration $n^*$. Each line represents a pair-interaction potential, where dashed blue line is the Coulomb interaction, full black line is the $v_{\text{co}}$ potential and thick pink line is $v_{\text{c}}^{\text{hc}}$. The other parameters are: $l_{\rm B}=e^2/(4\pi \varepsilon_0 \varepsilon k_{\rm B} T)=7\,\rm{\AA}$ and $a=3$\,\rm{\AA}.}}
\label{Fig1}
\end{figure}
%%%%%%%%%%%%%%%%%%%%%%%%%%%%%%%%%%%%%%%

%%%%%%%%%
\section{\Romannum{2}. Linearization of the equations of motion}
%%%%%%%%%
In this section we derive Eqs.~(17), (18) and~(21) of the Letter. The equations of motion can be linearized by writing $n_{\pm}({\boldsymbol r})  =n+\delta n_{\pm}({\boldsymbol r})$, $\rho ({\boldsymbol r})  =\delta\rho({\boldsymbol r})$, $\boldsymbol{u}({\boldsymbol r})  =\delta\boldsymbol{u}({\boldsymbol r})$ and $p({\boldsymbol r})  =p_0 + \delta{p}({\boldsymbol r})$,
while keeping only terms up to linear order in $\delta n_{\pm}$, $\delta\rho$, $\delta \boldsymbol{u}$, $\delta p$, and $\zeta_{\pm}$,
\beqa
\frac{\partial\delta n_{\alpha}}{\partial t} =&&  D_{\alpha}\nabla^{2}\delta n_{\alpha}-\mu_{\alpha}e_{\alpha}\boldsymbol{\nabla}\delta n_{\alpha}\cdot\boldsymbol{E}_{0}+\boldsymbol{\nabla}\cdot\sqrt{2D_{\alpha}n}\boldsymbol{\zeta}_{\alpha}  +\mu_{\alpha}ns_{\alpha}\int{\rm d}^{3}r'\delta\rho\left(\boldsymbol{r}'\right)\nabla^{2}V_{\text{co}}\left(\left|\boldsymbol{r}-\boldsymbol{r}'\right|\right)\\
 \eta\nabla^{2}\delta\boldsymbol{u}= && \boldsymbol{\nabla}\delta p-e\delta\rho\boldsymbol{E}_{0} \nn\\
 {\boldsymbol \nabla} \cdot  \delta  {\boldsymbol u} = &&0.\nn
\eeqa
In Fourier space, these equations become,
\beqa \label{Continuity_Fourier}
\frac{\partial\delta\tilde{n}_{\alpha}(\boldsymbol{k})}{\partial t}= && -D_{\alpha}k^{2}\delta\tilde{n}_{\alpha}(\boldsymbol{k})-i\mu_{\alpha}e_{\alpha}\delta\tilde{n}_{\alpha}(\boldsymbol{k})\boldsymbol{k}\cdot\boldsymbol{E}_{0}+ik\sqrt{2D_{\alpha}n}\tilde{\zeta}_{\alpha}(\boldsymbol{k})-\mu_{\alpha}ns_{\alpha}k^{2}\delta\tilde{\rho}(\boldsymbol{k})\tilde{V}_{\text{co}}(\boldsymbol{k})\nn \\
 \eta k^{2}\delta\boldsymbol{\tilde{u}}(\boldsymbol{k})= && -i\boldsymbol{k}\delta\tilde{p}(\boldsymbol{k})+e\boldsymbol{E}_{0}\delta\tilde{\rho}(\boldsymbol{k})\\
 \boldsymbol{k}\cdot\delta\boldsymbol{\tilde{u}}(\boldsymbol{k})= && 0,\nn
 \eeqa
where $s_{\pm}=\pm 1$, the tilde variables are $\tilde{f}(\boldsymbol{k})=\int {\rm d}^3r f(\boldsymbol{r}){\rm e}^{-i\boldsymbol{k}\cdot\boldsymbol{r}}$, and $\tilde{\zeta}_{\alpha}(\boldsymbol{k})$ is a scalar white noise function, satisfying
\beqa
\langle\tilde{\zeta}_{\alpha}(\boldsymbol{k})\rangle && =0\\
\langle\tilde{\zeta}_{\alpha}(\boldsymbol{k})\tilde{\zeta}_{\beta}(\boldsymbol{k'})\rangle && =\left(2\pi\right)^{3}\delta_{\alpha \beta}\delta\left(t-t'\right)\delta\left(\boldsymbol{k}+\boldsymbol{k}'\right),\nn
\eeqa
with $\alpha,\beta=\pm$. In order to obtain Eq.~(\ref{Continuity_Fourier}), we used the fact that $\boldsymbol{k}\cdot\boldsymbol{\tilde{\zeta}}_{\alpha}(\boldsymbol{k})=\sum\limits_{\alpha=1}^{3} k_{\alpha} \tilde{\zeta}^{i}_{\alpha}(\boldsymbol{k})$, is a sum of three independent white noise functions with zero mean. Therefore, it can be replaced by a single white noise function, whose variance is the sum of the variances of the three functions, $k\tilde{\zeta}_{\alpha}(\boldsymbol{k})$. Finally, Eq.~(\ref{Continuity_Fourier}) in matrix form gives Eqs.~(17) and (18) in the Letter.
\\[3pt]
\indent
Using the incompressibility condition to eliminate $\tilde{p}(\boldsymbol{k})$, we obtain $\delta\tilde{u}_{x}(\boldsymbol{k})$ in terms of $\delta\tilde{\rho}(\boldsymbol{k})$,
\beq
\delta\tilde{u}_{x}(\boldsymbol{k})=\frac{eE_{0}}{\eta}\frac{1}{k^{2}}\left(1-\frac{k_{x}^{2}}{k^{2}}\right)\delta\tilde{\rho}(\boldsymbol{k}).
\eeq
The two conductivity correction terms, $\kappa_{\text{hyd}}$ and $\kappa_{\text{el}}$, can be written as
\beqa \label{hyd_k}
\kappa_{\text{hyd}} && =\frac{e}{E_{0}}\langle\delta u_{x}\delta\rho\left(\boldsymbol{r}\right)\rangle\nn\\
 && =\frac{e^{2}}{\eta}\frac{1}{\left(2\pi\right)^{6}}\int{\rm d}^{3}k\int{\rm d}^{3}k'\:\frac{1}{k^{2}}\left(1-\frac{k_{x}^{2}}{k^{2}}\right) \langle\delta\tilde{\rho}(\boldsymbol{k})\delta\tilde{\rho}\left(\boldsymbol{k}'\right)\rangle{\rm e}^{i\left(\boldsymbol{k}+\boldsymbol{k}'\right)\cdot\boldsymbol{r}}
\eeqa
and
\beqa \label{ele_k}
\kappa_{\text{el}} && =-\sum_{\alpha}\frac{e\mu_{\alpha}}{E_{0}}\bigg\langle\int{\rm d}^{3}r'\delta n_{\alpha}\left(\boldsymbol{r}\right)\delta\rho\left(\boldsymbol{r}'\right)\partial_{x}V_{\text{co}}\left(\left|\boldsymbol{r}-\boldsymbol{r}'\right|\right)\bigg\rangle\nn\\
 && =-\sum_{\alpha}\frac{e\mu_{\alpha}}{ E_{0}}\frac{1}{\left(2\pi\right)^{6}}\int{\rm d}^{3}k\int{\rm d}^{3}k' (ik_{x})'\tilde{V}_{\text{co}}({k'})\langle\delta\tilde{n}_{\alpha}(\boldsymbol{k})\delta\tilde{\rho}(\boldsymbol{k}')\rangle{\rm e}^{i\left(\boldsymbol{k}+\boldsymbol{k}'\right)\cdot\boldsymbol{r}}.
\eeqa
Substituting the relation $\langle\delta\tilde{n}_{\alpha}(\boldsymbol{k})\delta\tilde{n}_{\beta}(\boldsymbol{k'})\rangle=\left(2\pi\right)^{3}C_{\alpha \beta}(\boldsymbol{k})\delta\left(\boldsymbol{k}+\boldsymbol{k}'\right)$ we obtain,
\beqa
\kappa_{{\rm hyd}} && =\frac{1}{\left(2\pi\right)^{3}}\frac{e^{2}}{\eta}\sum_{\alpha \beta}s_{\alpha}s_{\beta}\int{\rm d}^{3}k\frac{1}{k^{2}}\left(1-\frac{k_{x}^{2}}{k^{2}}\right)C_{\alpha \beta}(\boldsymbol{k}) \\
\kappa_{{\rm el}} && =\frac{1}{\left(2\pi\right)^{3}}\frac{e}{E_{0}}\sum_{\alpha \beta}\mu_{\alpha}s_{\beta}\int{\rm d}^{3}k\:(ik_{x})\tilde{V}_{\text{co}}({k})C_{\alpha \beta}(\boldsymbol{k}). \nn
\eeqa
Finally, the expressions for $\kappa_{\text{hyd}}$ and $\kappa_{\text{el}}$ can be further simplified employing the relations $C_{+-}(\boldsymbol{k})=C_{-+}^{*}(\boldsymbol{k})$ where $*$ is the complex conjugate, $C_{++}(\boldsymbol{k})=C_{--}(\boldsymbol{k})$, and using the fact that $C_{++}(\boldsymbol{k})$ and $\text{Re}\left[C_{+-}(\boldsymbol{k})\right]$ are even functions of $\boldsymbol{k}$ (see the next section), yielding Eq.~(21) of the Letter,
\beqa \label{integrals}
\kappa_{{\rm hyd}}&&=\frac{2{e}^{2}}{\eta}\int\frac{{\rm d}^{3}k}{\left(2\pi\right)^{3}}\frac{1}{k^{2}}\left(1-\frac{k_{x}^{2}}{k^{2}}\right)\big(C_{++}(\boldsymbol{k})-\text{Re}\left[C_{+-}(\boldsymbol{k})\right]\big)
\nonumber \\
\kappa_{{\rm el}}&&=\frac{2\bar{\mu}e}{E_{0}}\int\frac{{\rm d}^{3}k}{\left(2\pi\right)^{3}}\:k_{x}\tilde{V}_{\text{co}}({k})\text{Im}\left[C_{+-}(\boldsymbol{k})\right].
\eeqa
where $\bar{\mu}=(\mu_+ + \mu_-)/2$ is the average mobility.

%%%%%%%%%
\section{\Romannum{3}. Subtraction of the self-interaction}
%%%%%%%%%
We find the self-interaction term by re-writing $n_{\alpha}$ in the following way
\beq
n_{\alpha}(\boldsymbol{r})=\sum_{l}\delta\left(\boldsymbol{r}-\boldsymbol{r}_{l}\right)
\eeq
where the index $l$ sums over all ions of species $\alpha$, located at position $\boldsymbol{r}_{l}$. Then,
\beqa
\langle n_{\alpha}(\boldsymbol{r})n_{\alpha}(\boldsymbol{r})\rangle && =\sum_{l}\sum_{m}\langle\delta\left(\boldsymbol{r}-\boldsymbol{r}_{l}\right)\delta\left(\boldsymbol{r}-\boldsymbol{r}_{m}\right)\rangle.\\
\eeqa
The self-interaction part of $\langle n_{\alpha}\left(\boldsymbol{r}\right)n_{\alpha}\left(\boldsymbol{r}\right)\rangle$ is the sum of the terms for which $l=m$,
\beqa
\langle n_{\alpha}\left(\boldsymbol{r}\right)n_{\alpha}\left(\boldsymbol{r}\right)\rangle_{\rm s.i.} && =\sum_{l}\langle\delta\left(\boldsymbol{r}-\boldsymbol{r}_{l}\right)^{2}\rangle\nn\\
 && =\sum_{l}\frac{1}{\left(2\pi\right)^{6}}\int{\rm d}^{3}k\int{\rm d}^{3}k'\langle {\rm e}^{i\left(\boldsymbol{k}+\boldsymbol{k}'\right)\cdot\left(\boldsymbol{r}-\boldsymbol{r}_{l}\right)}\rangle\nn\\
 && =\sum_{l}\frac{1}{\left(2\pi\right)^{6}}\int{\rm d}^{3}k\int{\rm d}^{3}k'\langle {\rm e}^{i\boldsymbol{k}'\cdot\left(\boldsymbol{r}-\boldsymbol{r}_{l}\right)}\rangle\nn\\
 && =\frac{1}{\left(2\pi\right)^{3}}\int{\rm d}^{3}k\:\bigg\langle\sum_{l}\delta\left(\boldsymbol{r}-\boldsymbol{r}_{l}\right)\bigg\rangle\\
 && =\frac{1}{\left(2\pi\right)^{3}}\int{\rm d}^{3}k\:n.\nn\\
\eeqa
On the other hand,
\beq
\langle n_{\alpha}\left(\boldsymbol{r}\right)n_{\alpha}\left(\boldsymbol{r}\right)\rangle=n^2 + \frac{1}{\left(2\pi\right)^{3}}\int {\rm d}^3 k\:  C_{\alpha \alpha}(\boldsymbol{k}).
\eeq
Thus, the normalized part of $\langle n_{\alpha}\left(\boldsymbol{r}\right)n_{\alpha}\left(\boldsymbol{r}\right)\rangle$ (excluding self-interactions) is
\beqa
\langle n_{\alpha}(\boldsymbol{r})n_{\alpha}(\boldsymbol{r})\rangle_{{\rm norm}} && =\langle n_{\alpha}(\boldsymbol{r})n_{\alpha}(\boldsymbol{r})\rangle-\langle n_{\alpha}(\boldsymbol{r})n_{\alpha}(\boldsymbol{r})\rangle_{{\rm s.i.}}\nn\\
 && =n^2 +\frac{1}{\left(2\pi\right)^{3}}\int {\rm d}^3k\:\left[C_{\alpha \alpha}(\boldsymbol{k})-n\right],
\eeqa
which gives the normalized correlation matrix,
\beq \label{norm}
C^k_{\alpha \beta}\to C^k_{\alpha \beta}-n \delta_{\alpha \beta}.
\eeq
The correlation matrix $C_{\alpha \beta}(\boldsymbol{k})$ is found by solving Eq.~(20) in the Letter. Normalizing it according to the above Eq.~(\ref{norm}), we obtain
\beqa
 C_{++}(\boldsymbol{k})=C_{--}(\boldsymbol{k})= -\frac{n\cos\left(ka\right)\left(\frac{1}{2}\cos\left(ka\right)+\left(\frac{eE_{0}\lambda_{{\rm D}}}{k_{\rm B}T}\right)^{2}\left(\frac{k_{x}}{k}\right)^2+\lambda_{{\rm D}}^{2}k^{2}\right)}{\big(\cos\left(ka\right)+2\lambda_{{\rm D}}^{2}k^{2}\big)\left(\cos\left(ka\right)+\left(\frac{eE_{0}\lambda_{{\rm D}}}{k_{\rm B}T}\right)^{2}\left(\frac{k_{x}}{k}\right)^2+\lambda_{{\rm D}}^{2}k^{2}\right)}\nn
\eeqa
and
\beqa
C_{+-}(\boldsymbol{k})=C_{-+}^*(\boldsymbol{k})=\frac{n\cos\left(ka\right)\left(\frac{1}{2}\cos\left(ka\right)-i\frac{eE_{0}\lambda_{{\rm D}}^{2}}{k_{\rm B}T}k_{x}+\lambda_{{\rm D}}^{2}k^{2}\right)}{\big(\cos\left(ka\right)+2\lambda_{{\rm D}}^{2}k^{2}\big)\left(\cos\left(ka\right)+\left(\frac{eE_{0}\lambda_{{\rm D}}}{k_{\rm B}T}\right)^{2}\left(\frac{k_{x}}{k}\right)^2+\lambda_{{\rm D}}^{2}k^{2}\right)}.\nn
\eeqa
where the Fourier transform of $V_{\text{co}}(r)$, ${\tilde{V}_{\text{co}}({k})=e^{2}\cos\left(ka\right)/(\varepsilon_{0}\varepsilon k^{2})}$ was substituted.
%%%%%%%%%
\section{\Romannum{4}. Physical parameters}
%%%%%%%%%
%In Fig.~2, which shows the conductivity for aqueous solutions at room temperature, the following parameters were used: $T=25^{\circ}$C, $\varepsilon=78.3$, $\eta=0.890 \,\rm{mPa\cdot s}$,
In the following Tables~\Romannum{1} and~\Romannum{2} we present the parameters used to plot Figs.~2-4 in the Letter. Note that the length scales in the expressions for $\kappa_{\text{hyd}}$ and $\kappa_{\text{el}}$ in the Letter are calculated from the parameters in the two tables via $l_{\rm B}=e^2/(4\pi \varepsilon_0 \varepsilon k_{\rm B} T)$, $\lambda_{\rm D}=(8\pi l_{\rm B} n)^{-1/2}$ and $r_{\rm s}=1/(6\pi\eta\bar{\mu})$ where $\bar{\mu}=\kappa_{0} /(2e^2 n)$.
\begin{table}[h] \label{tab1}
\begin{tabular}{||c c c c||}
 \hline
 \,\,\,\,\,\,\,\,\,\,\,\,\,\,\,\,\,\, & $r_+ \rm{[\AA]}$ \,\,\,\,\,\, & $r_- \rm{[\AA]}$ \,\,\,\,\,\, & $\kappa_0/n [\rm{cm^2 \cdot S\cdot mol^{-1}}]$\\ [0.5ex]
 \hline\hline
 NaCl & $1.02$ & 1.81 & 126.39\\
 \hline
 KBr & $1.38$ & 1.96 & 151.9\\
 \hline
 LiI & $0.76$ & 2.20 & 115.46\\
  \hline
 KCl & $1.38$ & 1.81 & 149.79\\
 \hline
 KF & $1.38$ & 1.33 & 128.88\\
 \hline
\end{tabular}
\caption{The cation and anion radii~\cite{Shannon1976}, and the conductivity in the vanishing concentration limit, $\kappa_0$, normalized by $n$~\cite{Lide_Book2} for different salts in water (S is the Siemens electric conductance unit). We use the ``Effective ionic radii" by Shanon with 6-coordinate, while other sets for the ionic radii give very similar results~\cite{Shannon1976}.}
\end{table}
\begin{table}[h] \label{tab2}
\begin{tabular}{||c c c c||}
 \hline
 $ T\,\rm{[C^{\circ}]}$\,\,\,\,\,\, & $\eta$ $\rm{[mPa\cdot s]}$\,\,\,\,\,\, & $\varepsilon$ \,\,\,\,\,\, & $\kappa/n |_{n=0.01\rm{[M]}} [\rm{cm^2 \cdot S\cdot mol^{-1}}]$ \\ [0.5ex]
 \hline\hline
 5 & $0.152$ & 85.76 & 89.1 \\
 \hline
 25 & $0.890$ & 78.3 & 140.8 \\
 \hline
 50 & $0.547$ & 69.91 & 212.3 \\
 \hline
\end{tabular}
\caption{The viscosity~\cite{Korson1969}, dielectric constant~\cite{Malmberg1956}, and conductivity of KCl normalized by ion concentration at $n=0.01$\,M~\cite{Lide_Book2}, for three different temperatures.}
\end{table}
%
%%%%%%%%%%%%%%%%%%%%%%%%%%%%%%%%%%%%%%%%%%%
\section{\Romannum{4}. Failure of the theory at very high concentrations}
%%%%%%%%%%%%%%%%%%%%%%%%%%%%%%%%%%%%%%%%%%%
Our numerical results for the conductivity, Eqs.~(23) and (24) in the Letter, are shown to agree well with experimental results up to concentrations of a few molars. At even higher concentrations, however, the theory breaks down and predicts a diverging conductivity. This can be seen by the fact that the integral expressions for $\kappa_{\rm hyd}$ and $\kappa_{\rm el}$ diverge when the denominators have real roots, which happens when $a/\lambda_{\rm D}\gtrsim2.79$, or equivalently when $n\gtrsim 0.31/(a^2 l_{\rm B})$. The threshold thus depends on the finite-size parameter $a$. For NaCl in water, for example, the theory breaks when $n\gtrsim 9$\,M (which is far above the saturation concentration, namely not really physical). The failure stems from using the modified cutoff potential, which becomes inaccurate at very high concentrations, and leads to unphysical long-range order~\cite{Adar2019_2}.
%%%%%%%%%%%%%%%%%%%%%%%%%%%%%%%%%%%%%%%%%%%

%%%%%%%%%%%%%%%%%%%%%%%%%%%%%%%%%%%%%%%%%%%
\section{\Romannum{4}. The sensitivity of the computed conductivity to the choice of ion size}
%%%%%%%%%%%%%%%%%%%%%%%%%%%%%%%%%%%%%%%%%%%
Our theory relies on the incorporation of the finite ion size in the equations of motions, which eliminates unphysical electrostatic attraction between oppositely charged ions. Commonly used radii are the crystallographic radius $a_{\rm crys}$ that measures the bare radius, and the hydrated radius $a_{\rm hyd}$ that takes into account the hydration shell formed by the water molecules~\cite{Nightingale1959}.

In Fig.~\ref{Fig2}, we show our theoretical predictions for the conductivity of NaCl using different choices of our finite size parameter, $a$. For choices that deviate within 20\% from the crystallographic measurements, the results are still in reasonable agreement with experiments. However, when using the hydrated radius, which is 2-3 times larger than the crystallographic radius, the theory is much less accurate, and breaks down at concentration above 1.5\,molars, as the unphysical diverging conductivity regime discussed in Sec.\,I sets in. This is understandable since the hydration radius is 	measured for single ions (in very dilute solutions). Thus, it does not apply to more concentrated solutions when the hydration shells overlap.
%%%%%%%%%%%%%%%%%%%%%%%%%%%%%%%%%%%%%%%%%%%
%Fig2
\begin{figure}
\includegraphics[width = 0.4 \columnwidth,draft=false]{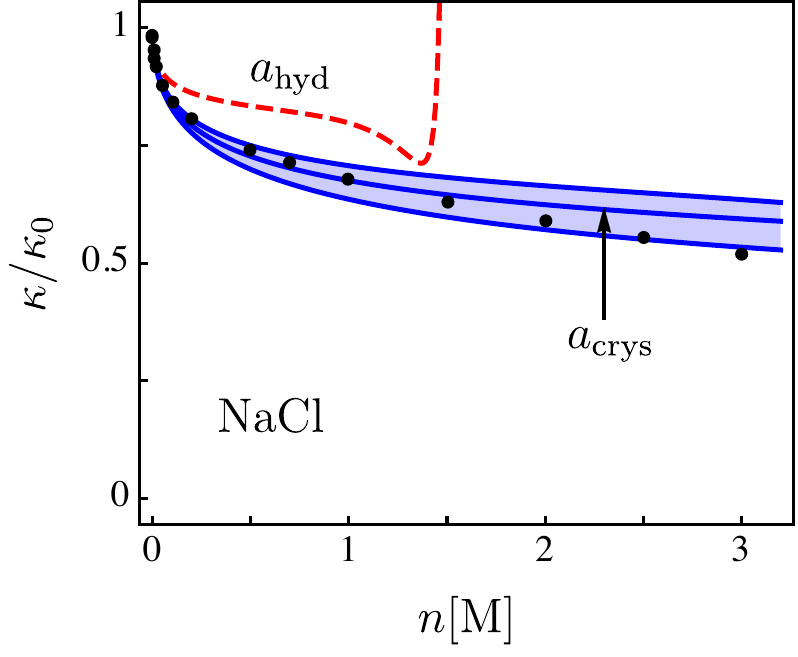}
\caption{\textsf{The conductivity, $\kappa$, of NaCl aqueous solution, normalized by $\kappa_0$, as a function of the salt concentration $n$. Blue lines from top to bottom are numerical results with $a=1.2a_{\rm crys},\,a_{\rm crys},\,0.8a_{\rm crys} $, respectively, where $a_{\rm crys}$ is the sum of the two crystallographic radii, $r_{\text{Na}}=1.02\,{\rm \AA}$, $r_{\text{Cl}}=1.81\,{\rm \AA}$. Red dashed line are numerical results with $a_{\rm hyd}$ that is the sum of the two hydrated radii taken from Ref.~\cite{Nightingale1959}, $r_{\text{Na}}=3.58\,{\rm \AA}$, $r_{\text{Cl}}=3.32\,{\rm \AA}$. Other parameters are as in Fig.~2 in the Letter.}}
\label{Fig2}
\end{figure}
%%%%%%%%%%%%%%%%%%%%%%%%%%%%%%%%%%%%%%%

 %%%%%%%%%%%%%%%%%%%%%%%%%%%%%%%%%%

%%%%%%%%

\end{document}